\documentclass[10pt,journal,compsoc]{IEEEtran}
\usepackage[T1]{fontenc}
\usepackage[utf8]{inputenc}
\usepackage{gensymb}
\usepackage{textcomp}
\usepackage{amsmath,amsthm,mathtools}
\usepackage{amssymb}
\usepackage{graphicx}
\usepackage[english]{babel}
\usepackage[normalem]{ulem}
\usepackage{url}
\usepackage{algorithm}
\usepackage{algorithmic}
\usepackage{cancel}
\usepackage{svg}

\usepackage{color}
\definecolor{blue}{rgb}{0,0,1}

\definecolor{dgreen}{rgb}{0,0.5,0}

\definecolor{dred}{rgb}{0.5,0,0}

\renewcommand\vec{\mathbf}
\renewcommand\d{\textrm{d}}
\usepackage{pbox} 
\usepackage{cite}

\begin{document}

\title{Predictability of Power Grid Frequency}

%

\author{Johannes~Kruse,
        Benjamin~Schäfer,
        Dirk~Witthaut
\IEEEcompsocitemizethanks{\IEEEcompsocthanksitem J. Kruse and D. Witthaut are with Forschungszentrum J\"ulich GmbH, Institute for Energy and Climate Research - Systems Analysis and Technology Evaluation (IEK-STE), 52428 J\"ulich, Germany, and also with the Institute for Theoretical Physics, University of Cologne, 50937 K\"oln,Germany. Email: \{jo.kruse, d.witthaut\}@fz-juelich.de \protect\\
\IEEEcompsocthanksitem B. Schäfer is with the School of Mathematical Sciences, Queen Mary University of London, London E1 4NS, United Kingdom. Email: b.schaefer@qmul.ac.uk}
\thanks{Manuscript received xxx; revised xxx; accepted xxx. Date of publication xxx; date of current version 27 Mar. 2020. (Corresponding author: Johannes Kruse)}}

\IEEEtitleabstractindextext{%
\begin{abstract}
The power grid frequency is the central observable in power system control, as it measures the balance of electrical supply and demand. A reliable frequency forecast can facilitate rapid control actions and may thus greatly improve power system stability. Here, we develop a weighted-nearest-neighbor (WNN) predictor to investigate how predictable the frequency trajectories are. Our forecasts for up to one hour are more precise than averaged daily profiles and could increase the efficiency of frequency control actions. Furthermore, we gain an increased understanding of the specific properties of different synchronous areas by interpreting the optimal prediction parameters (number of nearest neighbors, the prediction horizon, etc.) in terms of the physical system. Finally, prediction errors indicate the occurrence of exceptional external perturbations. Overall, we provide a diagnostics tool and an accurate predictor of the power grid frequency time series, allowing better understanding of the underlying dynamics. 
\end{abstract}

}

\maketitle
\IEEEpeerreviewmaketitle

\IEEEraisesectionheading{\section{Introduction}\label{sec:introduction}}
\IEEEPARstart{T}{he} electrical power system relies on a constant balance of supply and demand. Abundant energy will speed up generators and lead to an increase of the power grid's (mains) frequency. Similarly, a shortage of generation slows down the same generators and reduces the systems frequency as kinetic energy stored in the generator is transformed into electrical energy. Control systems, ordered from primary to tertiary control, help to ensure the balance of supply and demand by closely monitoring the frequency trajectory and maintaining it close to the desired reference value of $f=50$ or $60$ Hz \cite{Machowski2011}. Large deviations of the frequency away from the reference require decisive control actions and cause high costs \cite{wood2013power}.

To optimize the usage of costly control actions, we require a precise understanding of the power grid frequency. This frequency is neither constant nor varying slowly but is instead highly stochastic and subject to multiple external influences \cite{gorjao2020data, Anvari2019}. For example, the organization of the energy market leads to deterministic imprints of dispatch activities in the frequency in forms of sudden jumps or drops \cite{weissbach2009}. Simultaneously, an increasing share of renewable generators decreases the inertia available in the grid \cite{Milano2018} and introduces additional fluctuations \cite{Milan2013, Anvari2016}. 
Given this hybrid stochastic and deterministic nature, the question arises to which extend the frequency trajectory is predictable. A precise estimate of the future frequency trajectory would be very beneficial as it would allow an estimate of necessary control power early in time, saving costs \cite{wood2013power} and stabilizing the grid \cite{Machowski2011}.

Beyond precise forecasts of the near future trajectories, a fundamental understanding of the power grid frequency dynamics is critical as this one-dimensional time series encodes vast information on the stability and the current state of the power system \cite{dong_frequency_2014}. Only a solid understanding of how the energy mix, demand patterns and energy market rules impact the power system and its stability will allow us to implement and control highly renewable power systems in the future. As the starting point to develop such an understanding we study the power grid frequency since frequency data is much more readily available \cite{Schaefer2017a} than precise demand or generation values in a given synchronous area. 

With the increasing popularity of machine learning techniques \cite{alpaydin2009introduction}, there are many tools available to forecast time series, such as the power grid frequency. Recent studies used artificial neural networks (ANN)  \cite{kaur_power_2013} to predict hourly frequency time series in India based on features such as wind power generation and power demand. Other authors \cite{dong_frequency_2014} used a linear state space model and uncertain basis function to predict US frequency time series for up to one second, while a Bayesian network was used to predict the frequency time series for up to 3 minutes \cite{ma_analysis_2013}. Finally, ARMA processes have been used in the British grid to achieve prediction horizons of tens to hundreds of seconds \cite{bolzoni_real-time_2019} and in the US to achieve forecasts of 5 to 30 minutes \cite{bang_forecasting_2019}.

We will particularly focus on $k$-weighted-nearest-neighbor (WNN) methods, which have gained popularity in a variety of fields from biology \cite{kollmorgen2020nearest} to financial systems \cite{barkoulas2020nearest}, but have also been applied in the energy sector, e.g. to forecast electricity prices \cite{lora2007electricity} or power demand \cite{alvarez2010energy}.
In contrast to earlier applications of WNN predictors on the power grid frequency \cite{bang_forecasting_2019}, we improve and optimize the methodology, extend the prediction accuracy and horizon, compare different synchronous areas, and establish a system-relevant null model. Furthermore, we establish the accuracy and specificity of the forecasting algorithm as important characteristics to analyze the dynamics of the power system in general and the interplay of internal and external influences in particular. WNN predictors are particularly well suited to be interpreted as they are among the best explainable machine learning algorithms.

In this article, we use frequency recordings from several European synchronous areas to motivate the mean frequency (daily profile) as a suitable null model (Section \ref{sec:DataSource}) and develop a WNN predictor to forecast the time series (Section \ref{sec:Methods}). We demonstrate how our predictions outperform the null models in particularly on short prediction horizons and provide in-depth analysis and interpretation of when and how the power grid frequency can be predicted (Section \ref{sec:Results}).

\section{Dataset Description}
\label{sec:DataSource}

\begin{figure*}
    \centering
    \includegraphics[width=0.8\textwidth]{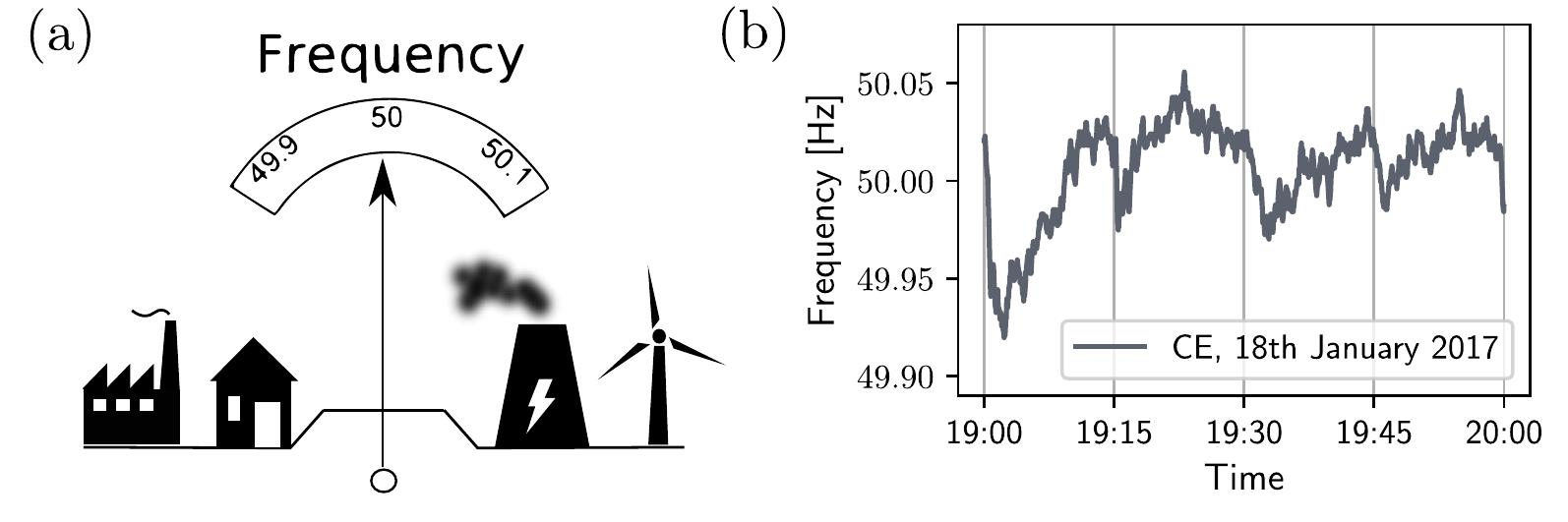}
    \caption{The nature of the power grid frequency. (a): The frequency reflects the balance of power demand and generation. Over-production causes a frequency increase and under-production a frequency decline. (b): Example frequency time series from the CE synchronous area \cite{noauthor_regelenergie_nodate}. It displays the typical frequency jumps at 15 minute intervals that are caused by the trading on electricity markets and subsequent changes of the power plant dispatch.}
    \label{fig:1_frequency_sketch_example}
\end{figure*}{}

\begin{figure*}
    \centering
    \includegraphics[width=\textwidth]{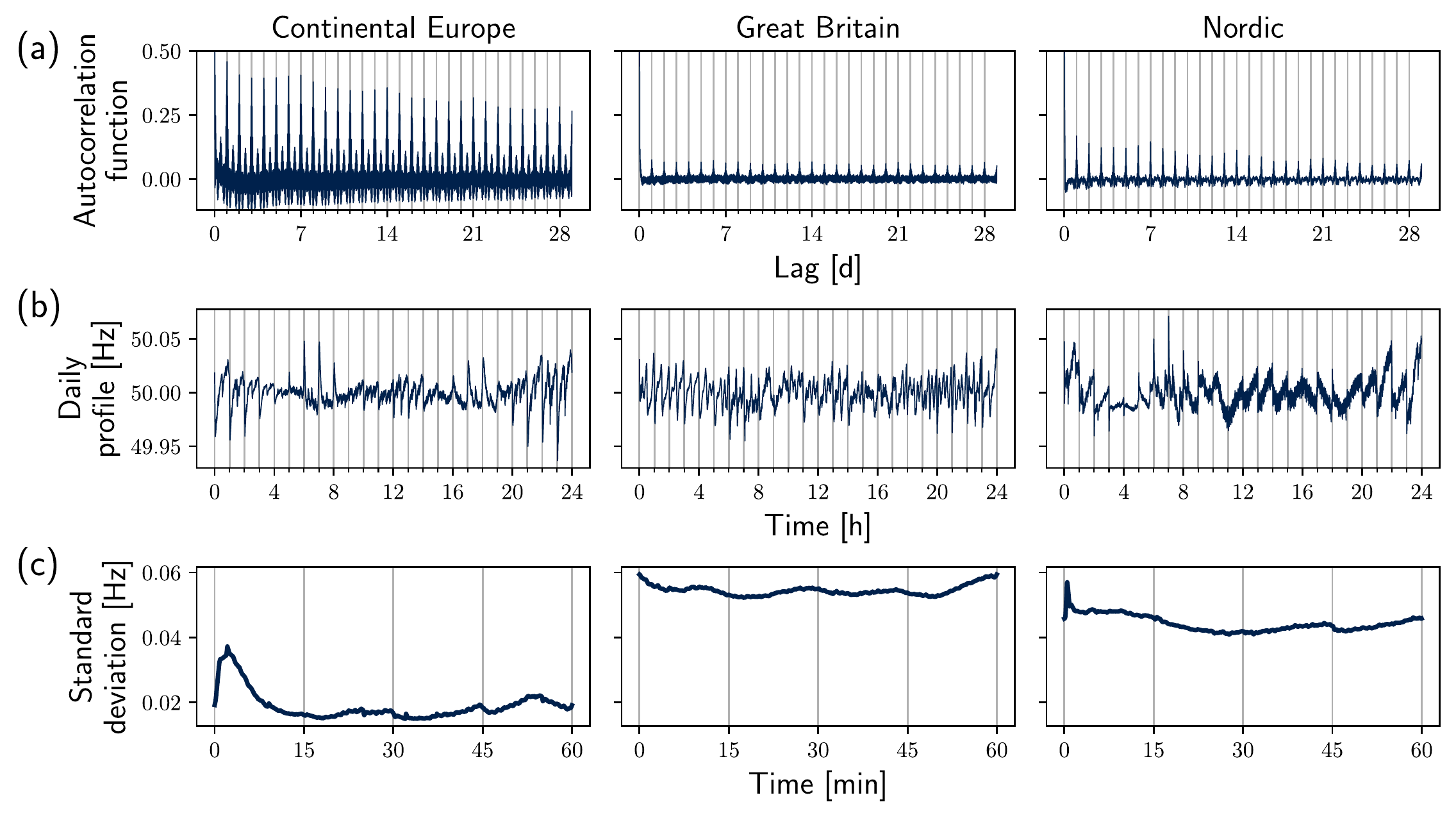}
    \caption{The daily profile is an important null model. (a): The autocorrelation functions show significant peaks that repeatedly occur with a period of 24 h. The one-day period thus is the main recurrence period for frequency patterns. Note, that the upper limit of the y-axis has been reduced from 1 to 0.5 in order to make even small peaks visible. (b) The daily profile is the average daily pattern that recurs with a one-day (24h) period. It is most pronounced in CE, where quasi-deterministic trading and dispatch actions play and important role.  (c): The standard deviation measures the variability among all frequency samples (in the training set) at a fixed time within the hour. The larger CE area displays the lowest variability, with a clear maximum at the beginning of the hour.}
    \label{fig:2_Mean_Standard_Deviation}
\end{figure*}{}

\subsection{Data Sources and pre-processing}
We train and test our frequency predictor on large high-resolution datasets from three different European synchronous areas. In particular, we use publicly available frequency recordings of the years 2015-2018 from the Continental Europe (CE) \cite{noauthor_regelenergie_nodate}, the Great Britain (GB) \cite{noauthor_historic_nodate} and the Nordic synchronous areas \cite{noauthor_frequency_nodate}, following the naming convention used in \cite{noauthor_commission_2017}. The data from CE and from GB comes with a one-second resolution, while the Nordic data exhibits a resolution of 0.1 s. Moreover, some of the datasets have varying formats and multiple frequency recordings are corrupted or missing. We therefore resample the data to a common one-second resolution and conduct a thorough pre-processing (Supplemental Material). The pre-processed time series are available online \cite{data_repository}, thus providing a ready-to-use database to develop new methods for frequency analysis and prediction.  

We want to point out that our pre-processing involves the identification and exclusion of corrupted measurements. However, the $k$-nearest-neighbor method can cope with the resulting holes in the time series. Missing segments are simply ignored during the nearest neighbor search. This is a great advantage of the WNN predictor, as we can harness the full length of the dataset without manipulating it too much. 


\subsection{Characteristics of the frequency time series}
The frequency trajectory exhibits deterministic as well as stochastic characteristics, which can be attributed to different dynamics within the power system. Firstly, a frequency deviation generally reflects a mismatch of power generation and demand (Fig. \ref{fig:1_frequency_sketch_example}(a)). Such a mismatch occurs when the power generation does not match the expected demand curve. The demand itself evolves continuously and shows typical daily, weekly and seasonal patterns \cite{wood2013power}. In contrast, the power generation exhibits discontinuous behaviour due to the trading on electricity markets and the resulting changes of the power plant dispatch \cite{weissbach2009}. In Europe, this trading is operated on various different spot-markets such as the European Energy Exchange Power Spot (EPEX SPOT), which covers countries in Western and Central Europe. The resulting dispatch changes are commonly scheduled for discrete time intervals of one hour, 30 and 15 minutes \cite{lin_electricity_2017,union_metis_2019}. The mismatch between the step-like behaviour of the generation and the continuous behaviour of the load leads to regular frequency jumps at the beginning of these trading intervals \cite{weissbach2009, Schaefer2017a, gorjao2020data}. Fig. \ref{fig:1_frequency_sketch_example}(b) shows a frequency sample that displays these typical quasi-deterministic jumps after every 15 minute interval. 

Secondly, the frequency characteristics are determined by the frequency control schemes. 
To assure a secure power system operation, these control measures drive back the frequency after a deviation from its reference value of 50 Hz \cite{Machowski2011}. They thus lead to a characteristic behaviour after a frequency jump or sag, which can for example be observed in Fig. \ref{fig:1_frequency_sketch_example}(b). On time scales of seconds after a disturbance, the inertia of the rotating generators and the energy supplied by primary control limits the frequency deviation caused by the disturbance. Afterwards, on time scales of several minutes, secondary and tertiary control set in and restore the system to a state of stable operation at the reference frequency \cite{Machowski2011}. 


Finally, the frequency characteristics are influenced by other external factors, that are of rather stochastic nature. Fluctuations of the demand directly affect the power balance, where demand forecasting errors \cite{Carpaneto2008} and large social events \cite{chen2011analysis} can lead to significant unexpected frequency deviations. The variability of renewable energy sources causes additional frequency fluctuations due to its inherent intermittency \cite{Haehne2018} or due to generation forecasting errors \cite{foley_current_2012}. In summary, the frequency characteristics are thus determined by a complex mix of stochastic and quasi-deterministic processes. 

\subsection{Analysis of frequency patterns}
\label{sec:frequency_patterns}
Despite its complex characteristics, the power grid frequency still exhibits regular patterns with a specific recurrence period. We identify this period by searching for regular peaks in the auto-correlation function (ACF) with time lags of up to one month (Fig. \ref{fig:2_Mean_Standard_Deviation}(a)). In all grid areas, the ACF displays regular peaks with a period of one day. Significant (but less pronounced) peaks with a period of 12h only show up in the CE data. In CE and GB, the ACF also exhibits regular peaks with shorter periods of 15 min, 30 min and 1h, but these peaks are much smaller than the daily peak \cite{Anvari2019}. Frequency patterns are thus most strongly correlated with patterns that occur one or multiple 24h-periods later. We conclude, that the one-day period is the main recurrence time for frequency patterns within all three synchronous areas.  

The average pattern that belongs to this main recurrence period is the mean daily frequency evolution, which we call the daily profile. A formal definition of the daily profile is given later in  \eqref{eq:daily_profile}. The daily profiles of our three datasets exhibit some common feature but also important differences (Fig. \ref{fig:2_Mean_Standard_Deviation}(b)). All profiles show pronounced frequency jumps at the beginning of the full hour, which reflect the impact of the hourly trading interval. In particular, the CE profile displays sharp peaks of different heights, while the peaks in the GB profile are the least pronounced. The direction and height of the peaks in the CE profile are time-dependent and related to whether the demand curve is rising or falling \cite{weissbach2009}. These results are consistent with the ACFs in Fig. \ref{fig:2_Mean_Standard_Deviation}(a). There, we also observe the strongest correlation for the CE data and the lowest correlation for the GB data. The CE frequency is thus strongly determined by regular daily patterns, while the GB frequency only exhibits weak patterns within this period. 

The relevance of regular patterns for the frequency time series is further characterized by the standard deviation (StD) in Fig. \ref{fig:2_Mean_Standard_Deviation}(c). We calculate the StD for each time within the hour, i.e. the StD at 0 min is computed as the StD of all frequency recordings with time stamps $XX:00:00$ averaging over all hours $XX$ and days. In general, CE exhibits the lowest and GB the highest variability. The StD peaks after the full hour trading event in the Nordic and especially in CE areas, where the StD almost doubles after the full hour trading peak. The exact value of the full hour frequency peak thus exhibits a particularly high uncertainty.

We conclude, that CE is a comparatively low-noise system with defining deterministic events that drive the standard deviation. Deterministic patterns are least pronounced in GB, such that random fluctuations are of highest importance compared to the other areas. The Nordic data is mostly in between. The differences between the grid areas can be attributed to different system properties as well as varying regulations for frequency control and market operation. For example, the low variance in the CE area is likely related to its large size \cite{Schaefer2017a}, which provides much inertia and enables spatial balancing of nodal power mismatches. Moreover, the deadband, i.e. the frequency range without active control, is the largest in GB thus resulting in a high frequency variability \cite{noauthor_commission_2017}. Despite these differences, there is one important common result: In all three cases the main recurrence period of frequency patterns is one day. The same result was found for frequency time series from US grids \cite{bang_forecasting_2019}. This highlights the importance of the daily time scale and the corresponding daily profile for the prediction of future frequency patterns.

\section{Forecasting Methods}
\label{sec:Methods}

\begin{figure*}
    \centering
    \includegraphics[width=\textwidth]{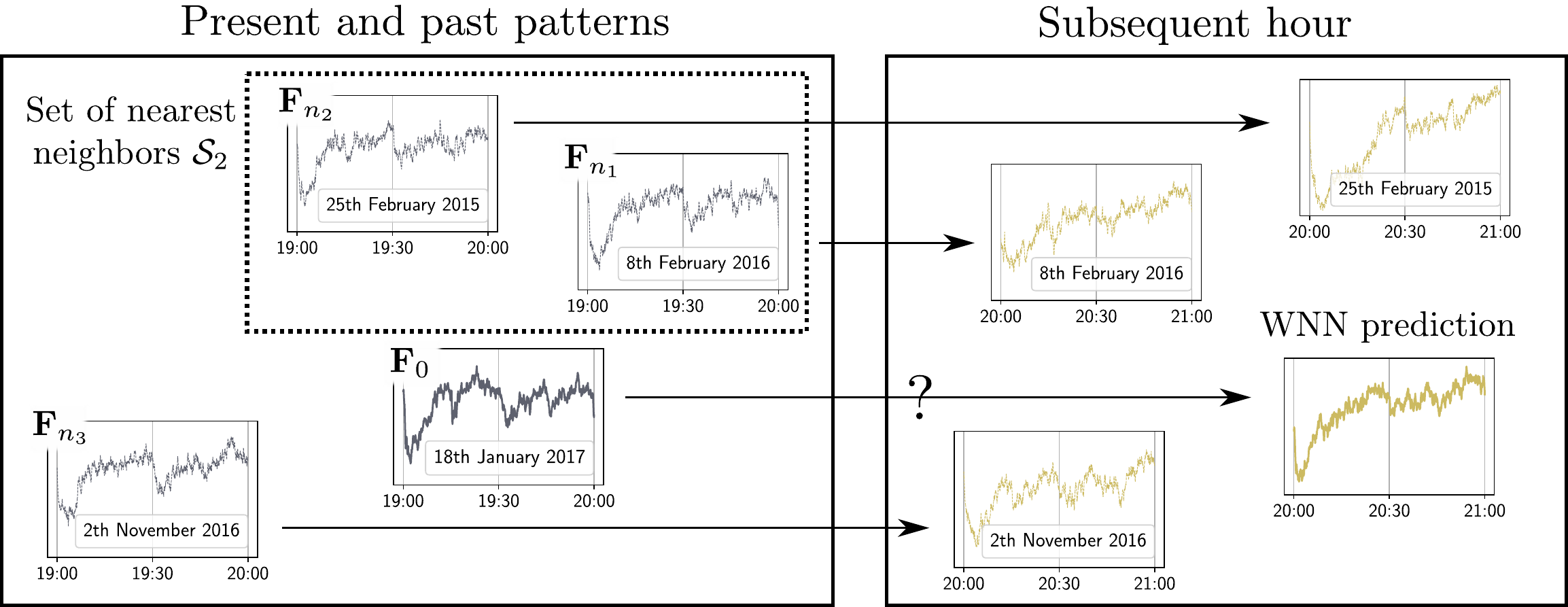}
    \caption{The WNN predictor searches for similar patterns in the past. To predict the future of the present (initial) pattern $\vec F_0$, the WNN method looks for similar patterns $\vec F_{n_j}$ in the past. The patterns that are most similar to the initial pattern form the set of nearest neighbors. Here, we have chosen a set $\mathcal S_2$ of two nearest neighbors. The average of their subsequent trajectories generates the WNN prediction.}
    \label{fig:3_wnn_sketch}
\end{figure*}{}

\subsection{Weighted Nearest neighbors}
The WNN method predicts future values of a time series by looking for similar patterns in the past. To predict the frequency $f(t)$ for $t\ge t_0$, we thus cut the historical time series into non-overlapping patterns $\vec F_n$ with $\gamma$ data points and a time delay $\tau$:
\begin{align}
    \vec F_n = \begin{pmatrix} f(t_0 -(n+1)\gamma\tau) \\ 
                f(t_0 - (n+1)\gamma\tau + \tau)\\ 
                f(t_0 - (n+1)\gamma\tau + 2\tau) \\
                \dots\\ 
                f(t_0 - n\gamma\tau - \tau)
            \end{pmatrix} .
\end{align}
The vectors $\vec F_n$ form an embedding of the time series in a space of dimension $\gamma$, which is also referred to as delay embedding in the context of time series analysis \cite[Chap. 2]{schelter_handbook_2006}. To include the information of all data points, we choose a delay equal to the original time resolution of $\tau=1$ s. 

The WNN predictor searches for patterns $\vec F_n$ that are similar to the initial pattern $\vec F_0$, which ends at the prediction start $t_0$. However, we already know that frequency patterns mainly recur with a period of one day (Section \ref{sec:DataSource}). Therefore, we only look for similar patterns at the same time of the day, i.e. only within the set
\begin{align}
    \mathcal F = \{ \vec F_n | \exists i \in \mathbb N: n\gamma\tau =  i\cdot 24 \textrm{h} \}.
    \label{eq:same_hour_patterns}
\end{align}
From this set, we choose those patterns that are closest to the initial pattern in terms of the distance
\begin{align}
    \d(\vec F_n) 
    &= \| \vec F_n - \vec F_0 \| \nonumber.
\end{align}
Here, $\| \cdot \|$ denotes the Euclidean distance. Given this metric, we sort the patterns as $\d(\vec F_{n_1}) \le \d(\vec F_{n_2}) \le ... \le \d(\vec F_{n_M})$, $M=|\mathcal F|$ being the total number of patterns. We then select $k$ patterns with the smallest distance to the initial pattern and obtain the ordered set of nearest neighbors
\begin{align}
    \mathcal S_k = \{ n_1, n_2,..., n_k | \vec F_{n_j} \in \mathcal F \}.
\end{align}
In practice, we use the \textit{scikit-learn} package to search and sort the nearest neighbors \cite{scikit-learn}.

Finally, we assume that trajectories, which were similar in the past, will likely be similar in the future
(Fig. \ref{fig:3_wnn_sketch}). Technically, the prediction $f_p(t_0 + \Delta t)$ is therefore given by a weighted average of the trajectories succeeding the $k$-nearest-neighbors:
\begin{align}
    f_p(t_0 + \Delta t) = \frac{1}{\sum_{j=1}^k \alpha_j} \sum_{j=1}^k  \alpha_j f(t_0 - n_j\gamma\tau + \Delta t ). \label{eq:prediction}
\end{align}
The weights $\alpha_n$ are chosen to decrease with the distance $\d(\vec F_{n_j})$ which introduces an additional smoothing \cite[Chap. 3]{schelter_handbook_2006}. Following \cite{lora2007electricity}, we use a linear weighting that has the following form:
\begin{align}
    \alpha_j = \frac{\d(\vec F_{n_k}) - \d(\vec F_{n_j})}{\d(\vec F_{n_k})-\d(\vec F_{n_1})}.
\end{align}
In practice, we apply the WNN method to predict the time steps $\Delta t \in \{1s, 2s, ..., T\}$ with a maximum prediction length of $T=3600$s. Longer predictions are not relevant, since the superiority of the WNN method over the null models is mostly revealed within the first 30 minutes of the prediction (see Section \ref{sec:Results}). 

\subsection{Performance estimation}
During the optimization and evaluation of the WNN predictor, we use the Mean Square Error (MSE) as the central performance measure. In particular, we evaluate the time-dependent MSE of a general predictor $\hat f(t_0 + \Delta t)$ for each prediction step $\Delta t$ by averaging over different starting times~$t_0^i$:
\begin{align}
     \textrm{MSE}_{\Delta t}(\hat f) = \frac{1}{N} \sum_{i=1}^N \left( \hat f(t_0^i+ \Delta t) - f(t_0^i + \Delta t)\right)^2 .
     \label{eq:mse_t}
\end{align}
To select different starting times, we randomly choose $N=5000$ different start hours $h_0^i$. The starting time is then given by $t_0^i = h_0^i + \Delta t_0$ where $h_0^i$ counts the hours after the start of 2015 and $\Delta t_0$ represents a fixed starting time within the hour. In this way, we account for the frequency dynamics, that crucially depend on the time within the hour as discussed in Section \ref{sec:DataSource}.  

To estimate the out-of-sample performance of our predictor, we split our data into different subsets (equally for all synchronous areas). In general, the years 2015 and 2016 serve as training set, which is searched for nearest neighbors during the WNN prediction. To optimize the hyperparameters of the WNN predictor, we evaluate its MSE on a validation set that comprises the year 2017 (Section \ref{sec:optimization}). Finally, we define the year 2018 as our test set. On the test set, we compare the performance of our WNN predictor to system-specific null models (Section \ref{sec:null-model}). 

\subsection{Hyperparameter optimization}
\label{sec:optimization}
Our WNN method exhibits two hyperparameters which are the number of nearest neighbors $k$ and the window size (or pattern length) $\gamma\tau$. We use a window size of $\gamma\tau=60~$min unless stated otherwise, which provides a good prediction at low computational effort. The window size is thus not explicitly optimized, but we investigate its impact on the prediction accuracy in Section \ref{sec:impact_window_size}. 

In contrast, we strictly optimize the number of nearest neighbors $k$ by using two different approaches. In the \textit{fixed-$k$} approach, we estimate an optimal number of nearest neighbors by minimizing the time-averaged prediction error MSE$(f_p)$ of the WNN predictor $f_p$: 
\begin{align}
      \textrm{MSE}(f_p) = \frac{1}{T} \sum_{\Delta t = 1s}^T \textrm{MSE}_{\Delta t}(f_p).
\end{align}
In practice, we perform a grid search on the set $\mathcal G =\{ 1,3,5,...,451 \}$ to determine a fixed optimal value $k_{opt} \in \mathcal G$ for all prediction times $\Delta t \in [1s,T]$. This is how the WNN method is commonly used \cite{lora2007electricity, alvarez2010energy}. We denote this as \textit{fixed-$k$} WNN prediction. 

In the \textit{adaptive-$k$} approach, we minimize the time-dependent error $MSE_{\Delta t}(f_p)$  \eqref{eq:mse_t} for each prediction step $\Delta t$ individually. In this way, we account for the very different prediction horizons we investigate in our paper. These range from several seconds to one hour thus making it highly probable to obtain different optimal $k$-values for different prediction horizons. In practice, we therefore calculate a time-dependent estimator $k_{opt}(\Delta t)$ for each prediction step $\Delta t$ by performing a grid search on the set $\mathcal G$. To make the estimator more robust against noise, we smooth $k_{opt}(\Delta t)$ using a sliding window with a length of one minute. Finally, the \textit{adaptive-$k$} WNN prediction is calculated by simply inserting a time-dependent $k$ into \eqref{eq:prediction}.

\subsection{Null models}
\label{sec:null-model}
On our test set, we compare different predictors based on their Root Mean Square Errors (RMSE), which reflects the actual frequency error in Hz:
\begin{align}
    \textrm{RMSE}(\hat f) = \sqrt{\textrm{MSE}_{\Delta t}(\hat f)}.
\end{align}
We use two easily interpretable null models to benchmark the performance of the WNN predictor. Our first trivial null model is the reference value of 50Hz, which is also the frequency mean:
\begin{align}
    f_m(t_0 + \Delta t) = 50 \, \textrm{Hz}.
\end{align}
Our second null model is the daily profile. In Section \ref{sec:DataSource}, we have shown that the daily profile is the most important system-specific pattern that recurs with a period of one day. It should therefore be a benchmark model for every newly proposed frequency predictor. In practice, we calculate the daily profile predictor by averaging over all the patterns in the set $\mathcal F$ (from  \eqref{eq:same_hour_patterns}):
\begin{align}
    f_d(t_0 + \Delta t) = \frac{1}{|\mathcal F|} \sum_{n \in \mathcal F} f(t_0 - n\gamma\tau + \Delta t).
    \label{eq:daily_profile}
\end{align}{}
To make its prediction comparable to the WNN predictor, we have restricted the set $\mathcal F$ to patterns from the training set. Note, that the WNN predictor \eqref{eq:prediction} converges to the daily profile predictor in the limit $k \rightarrow \infty$ when applying uniform weights. 

\section{Results}
\label{sec:Results}

\subsection{Forecast examples}

\begin{figure*}
    \centering
    \includegraphics[width=\textwidth]{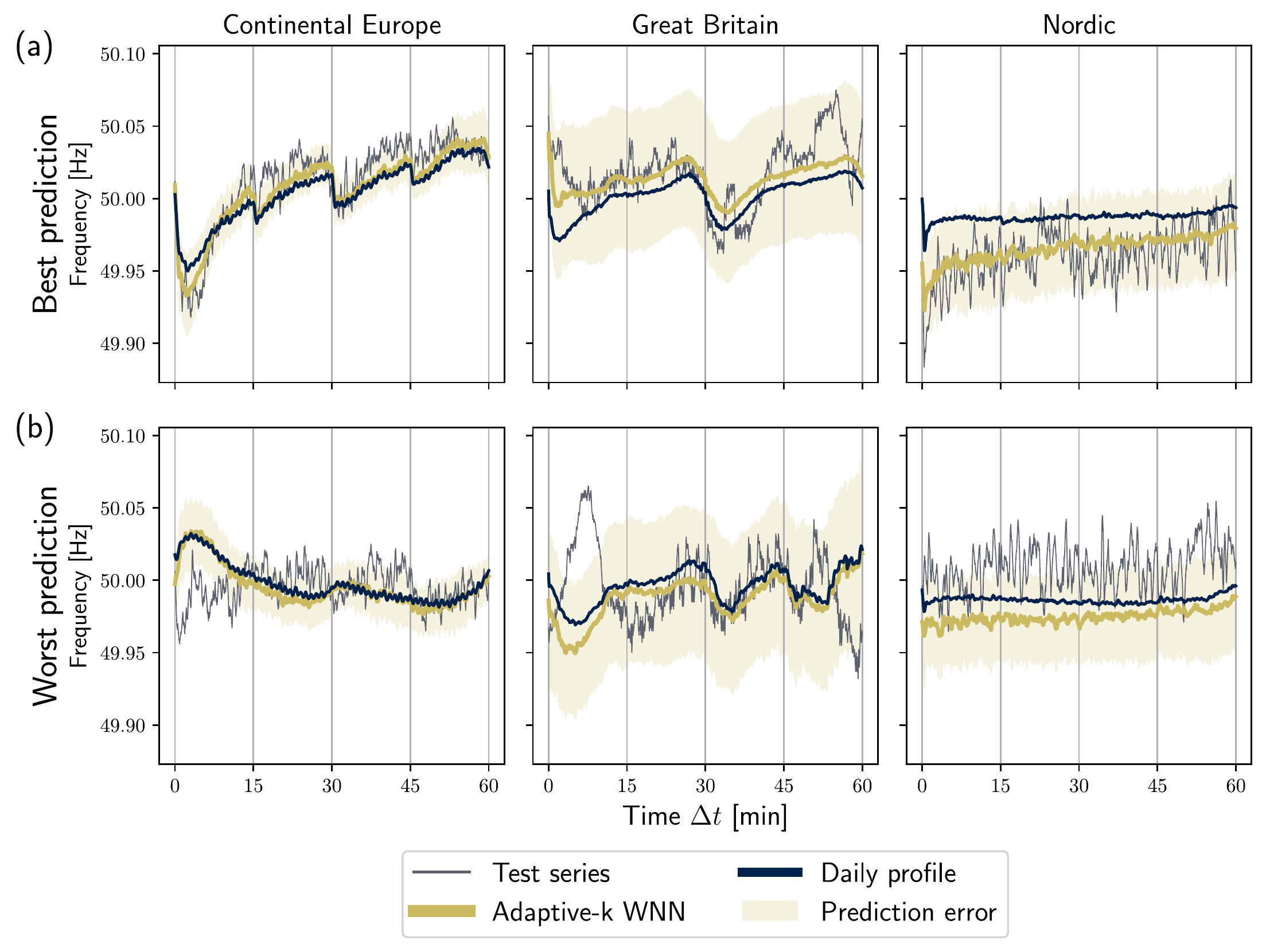}
    \caption{The best predictions are a smoothed version of the real trajectory. Here, we present the best (a) and worst (b) adaptive-$k$ predictions from the test set. The selection is based on the relative error RMSE($f_p$)/RMSE(50Hz). With that we account for the difference in variance among the samples, that would automatically result in higher or lower error values. The prediction error $\sigma_{\Delta t}$ \eqref{eq:uncertainty_estimate} equals one standard deviation within the largest set of nearest neighbors used during the prediction. It is thus an upper bound for the standard deviation of the adaptive-$k$ WNN prediction.}
    \label{fig:4_Best_worst_Predictions}
\end{figure*}{}

The best and worst prediction examples give us a first impression about the performance of the WNN predictor (Fig. \ref{fig:4_Best_worst_Predictions}). We complement these examples with
an estimate of the prediction uncertainty $\sigma_{\Delta t}$ that is based on the StD of the nearest neighbors:
\begin{align}
    \sigma_{\Delta t}^2 = \langle f(t_0 - n \gamma \tau + \Delta t)^2 \rangle  -\langle f(t_0 - n \gamma \tau + \Delta t) \rangle^2 \label{eq:uncertainty_estimate}
\end{align}
Here, $\langle \cdot \rangle$ denotes the average over all $n \in \mathcal S_k$. For the adaptive-$k$ WNN, we use $k= \max_{\Delta t} k_{opt}(\Delta t)$ , which turns  \eqref{eq:uncertainty_estimate} into an upper bound for the uncertainty. 

We observe, that the best predictions are essentially a smoothed curve of the real trajectory. The prediction is often very similar to the daily profile, but performs better especially in the first 15 minutes. Even more, the prediction uncertainty provides a good estimate for the short-term variability of the real data. 

The worst predictions in GB and CE make mistakes at the boundaries but still capture the remaining trajectory (e.g. 30-45 min in GB).  In both examples, the daily profile and the WNN forecast predict the same direction for the hourly frequency jump but the real trajectory deviates in the opposite direction. The deviation indicates unforeseen events affecting the grid frequency trajectory, which are also not captured at all by the daily profile. This relation points to a potential application of time series prediction in the posteriori analysis of power system operation. A large forecasting error can serve as a tool to identify external (unforeseen) events. 

Meanwhile, the worst prediction in the Nordic area stays nearly constant and the corresponding real trajectory randomly oscillates around a shifted value. This exemplifies the weak performance of the WNN predictor for unspecific patterns with strong noise. 

\subsection{Performance of forecasting methods}
\label{sec:performance_eval}

\begin{figure*}
    \centering
    \includegraphics[width=\textwidth]{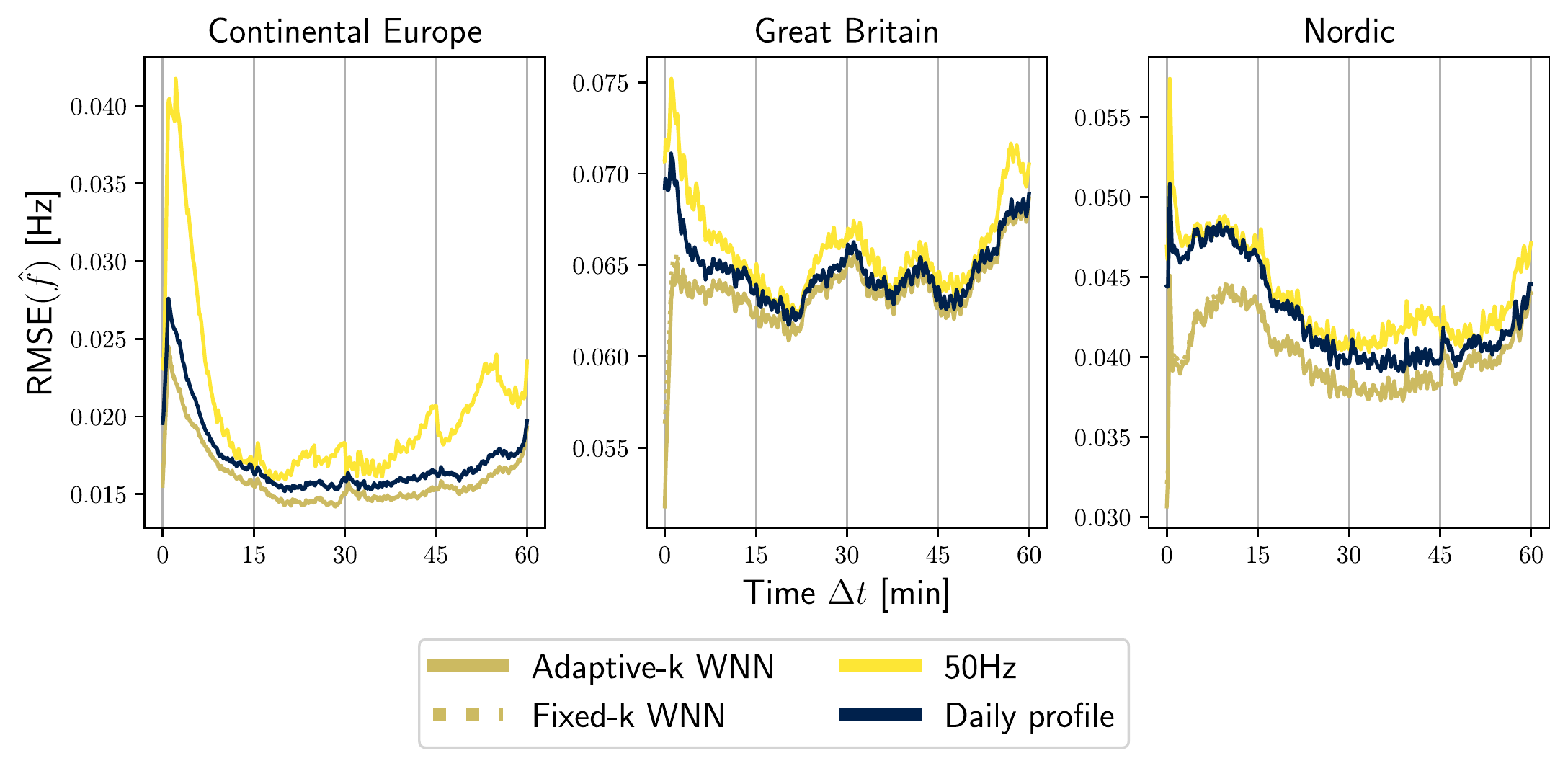}
    \caption{The WNN predictors outperform alternatives. The WNN predictor outperforms the null models in all three synchronous areas by returning the smallest RMSE, especially in the first 15-30 minutes. The scale of the y-axis differs between the subplots, since the GB area exhibits much larger errors than the CE area. The RMSE of the WNN predictor is further strongly time-dependent and converges to the daily profile towards the end of the prediction. }
    \label{fig:5_k-NN_vs_50Hz_vs_Daily}
\end{figure*}{}

We evaluate the performance of our forecasting methods by calculating their RMSE on our test set (Fig. \ref{fig:5_k-NN_vs_50Hz_vs_Daily}). The results show that our WNN predictor outperforms both null models in all grid areas. Its RMSE is smallest for CE and largest for GB. This relates to Section \ref{sec:DataSource} where we identified GB as the most stochastic and CE as the most deterministic and thus most predictable grid. The improvement of the WNN predictor relative to the daily profile is largest in Nordic (up to 30\%) and smallest in CE (up to 20\%). This is due to the fact, that the daily profile itself is already a good predictor in CE. Meanwhile, the daily profile performs much worse in the Nordic area, where its RMSE nearly follows the 50Hz prediction error.

Comparing performance over time, we observe that the WNN outperforms the null models especially during the first 15min.  As the prediction length increases, the WNN prediction converges to the daily profile. On the other hand, the performance is also clearly affected by the trading events (especially in CE). This time-dependence will be investigated in more detail in Section \ref{sec:start_time} and \ref{sec:impact_window_size}. 

Finally, we note that there is no significant difference between the adaptive-$k$ and the fixed-$k$ WNN predictor for long predictions of up to 60 minutes (Fig. \ref{fig:5_k-NN_vs_50Hz_vs_Daily}). However, we observe a significant difference for very short prediction horizons, which we will discuss in the next section.


\subsection{Optimal number of nearest neighbors}
\label{sec:optimal_k}

\begin{figure*}
    \centering
    \includegraphics[width=\textwidth]{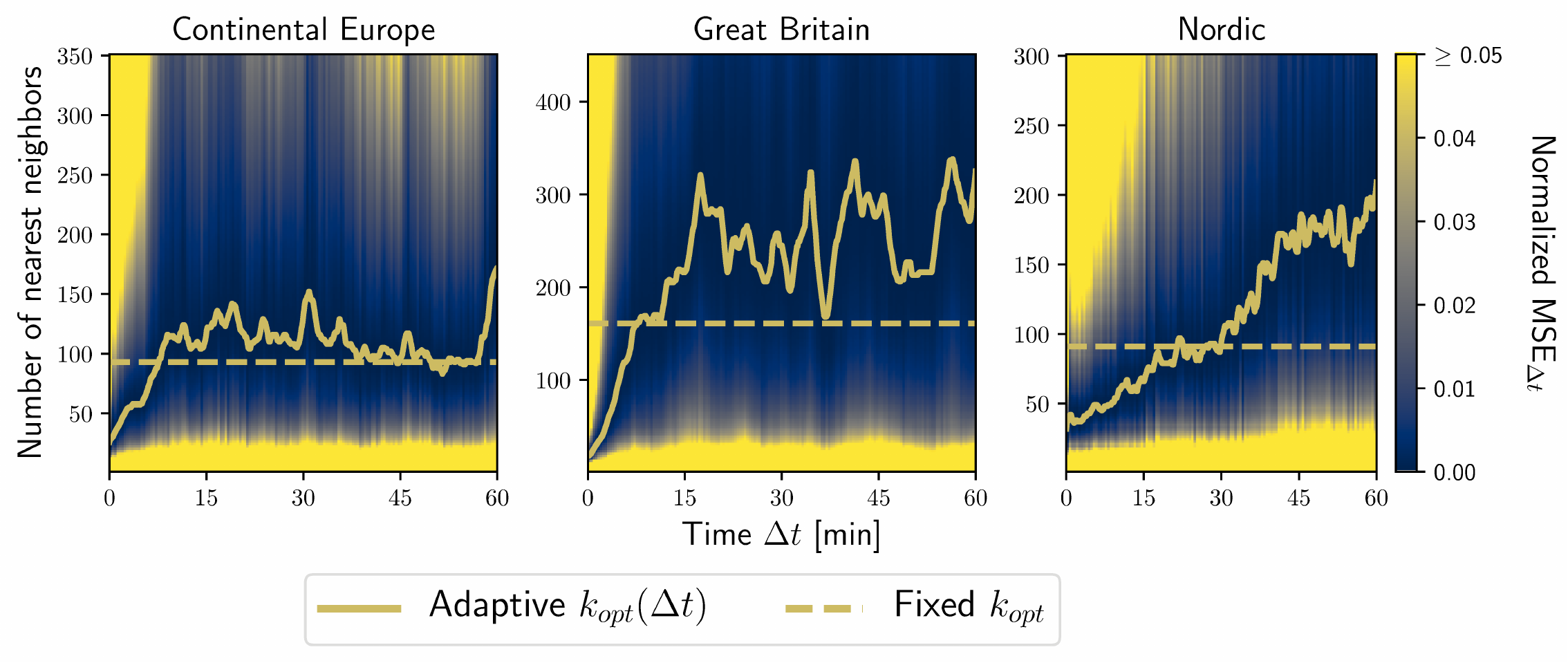}
    \caption{The Optimal number of nearest neighbors increases over time. To compare the error landscape for different time steps $\Delta t$, we normalize the MSE in this figure. The normalization rescales the MSE to values between zero and one for each time step $\Delta t$. The time-dependent minimum of this landscape is the adaptive number of nearest neighbors $k_{opt}(\Delta t)$. The fixed $k_{opt}$ minimizes the aggregated MSE leading to very similar prediction errors in all but the the first minutes.
    }
    \label{fig:6_Optimal_nr_neasrest_neighbors}
\end{figure*}{}

Determining the optimal number of nearest neighbors $k_{opt}(\Delta t)$ can help to better understand the functioning of the WNN predictor. Moreover, it yields valuable information about the grid frequency dynamics in general. We present the optimization results in Fig. \ref{fig:6_Optimal_nr_neasrest_neighbors}, which shows the normalized RMSE landscape as a function of $k$ and $\Delta t$ as well as the optimal values $k_{opt}(\Delta t)$. The adaptive number of nearest neighbors tends to increase the more the prediction is in the future. However, the minimum is very flat at most time steps and both the adaptive-$k$ and the fixed-$k$ predictor lead to very similar errors (in agreement with results from Section \ref{sec:performance_eval}). We only observe a significant difference within the first minute, where the adaptive-$k$ WNN yields up to 5\% better results than the fixed-$k$ approach. We conclude, that the adaptive approach is slightly better, especially in the first 1min. We will therefore only apply the adaptive-$k$ WNN method throughout the rest of the paper. 

As an application, we can interpret $k_{opt}(\Delta t)$ in terms of the predictability of frequency patterns.
A low number of nearest neighbors corresponds to well-defined trajectories that match to some past trajectories accurately. Contrary, a higher number of nearest neighbors $k_{opt}(\Delta t)$ indicates that trajectories are rather unspecific with respect to the history. A large number of trajectories has to be averaged such that the prediction is similar to the daily profile. In particular in the first 15 minutes, the adaptive-$k$ yields very low $k$ values. The frequency trajectory is thus very specific in this time regime. As the prediction time increases, the optimal number $k_{opt}(\Delta t)$ rises. The trajectory thus becomes more unspecific with respect to past patterns and thus less predictable for the WNN predictor. Consistently, the WNN predictor approaches the daily profile at the end of the hour, which we obtain for $k\rightarrow \infty$. 

\subsection{Impact of the prediction start}
\label{sec:start_time}

\begin{figure*}
    \centering
    \includegraphics[width=\textwidth]{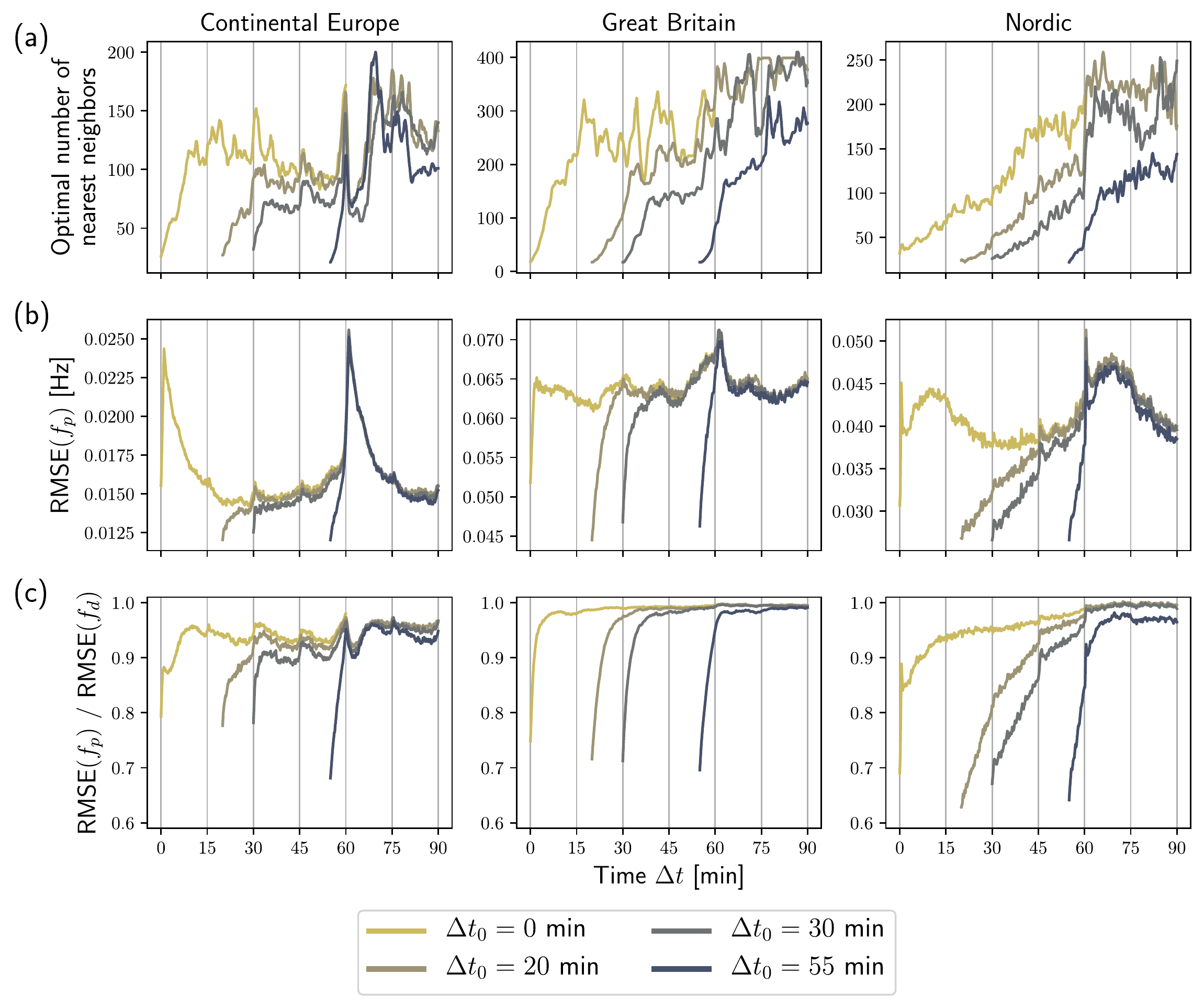}
    \caption{Trading events shorten the prediction horizon. Here, we show the optimal number of nearest neighbors $k_{opt}(\Delta t)$ (a), the RMSE (b) and the relative RMSE (c), which is normalized by the daily profile error. Irrespective of the starting time within an hour $\Delta t_0$, the predictions perform best within a time horizon of 15min. However, trading events introduce additional uncertainty thus increasing the prediction error and shortening the prediction horizon.}
    \label{fig:7_Starting_time}
\end{figure*}{}

Up to now we have focused on predictions starting at full hours, such that the prediction interval coincides exactly with the main time scale of energy trading and power plant dispatch. We now widen our scope and assess the time-dependence of the WNN performance by initializing the prediction at different starting times $\Delta t_0$ (Fig. \ref{fig:7_Starting_time}). To still relate the WNN performance to our null models, we additionally assess its relative error RMSE$(f_p)$/RMSE$(f_d)$ ("relative RMSE"), which is normalized by the daily profile error RMSE$(f_d)$.

Irrespective of the trading events, we observe two different time regimes depending on the prediction length. During the first 15 minutes, the relative RMSE and the optimal number $k_{opt}(\Delta t)$ are increasing while still being much lower than future values. Here, the frequency dynamics exhibit specific patterns, that resemble particular patterns in the past (as described in Section \ref{sec:optimal_k}). This specific memory is lost over time, as the relative RMSE increases continuously during the first 15-30 minutes. In particularly in the CE and Nordic areas, one can identify two clearly distinct time scales of memory loss: Firstly, there is an initial rapid increase of the RMSE and the relative RMSE within approximately one minute. It is followed by a slower, not necessarily monotonous increase of the relative RMSE on timescales up to tens of minutes. This clear separation of time scales is especially visible when energy trading is important, i.e. at full hours being strongest in the CE area. It could be attributed to the grid inertia or to control measures that provide additional memory for a short period of time. 

Finally after 15-30 minutes, the relative RMSE reaches a relatively constant level in CE and GB with values close to one. Here, the WNN prediction does not differ much from the daily profile anymore. Meanwhile, the relative RMSE and the optimal number $k_{opt}(\Delta t)$ continue to rise for up to 60 minutes in the Nordic area. Here, the memory of specific historic patterns thus reduces much slower compared to the other areas. We will come back to this effect in Section \ref{sec:impact_window_size}.

In addition to the prediction length, the trading events play a crucial role for the prediction. In all grid areas, the RMSE increases strongly around the one-hour trading event. For CE and Nordic, we observe this also at 15 and 45 minutes. Around these events, the dispatch is changed abruptly, causing large frequency deviations, which are hard to forecast accurately (Fig. \ref{fig:2_Mean_Standard_Deviation}(c)). The optimal number of nearest neighbors $k_{opt}(\Delta t)$ and the relative RMSE also peak at the trading event. This indicates a lack of specific information about the trading peak and a high uncertainty connected to it. CE is a special case, as its one-hour trading jump is particularly hard to forecast. Interestingly, $k_{opt}(\Delta t)$ decreases again after the peak. The trajectory thus becomes more specific and predictable again, probably due to the control measures reacting to the disturbance in a pre-defined way. 

The trading peaks have another important impact on the prediction error. After a trading event, the RMSE  loses its dependence on the starting time $\Delta t_0$ and joins the error curve of earlier prediction starts. This happens in all grid areas, at latest during the one-hour trading event. In practice, it means that our prediction starting at 55 min performs approximately as well at 60 min as the one that started at 0 min. The information contained in the initial pattern thus looses its significance with the occurrence of a trading event. In other words, the trading events cause a memory loss in the frequency trajectory. 

We conclude, that the best WNN prediction is always obtained right after the prediction starts. On a time horizon of up to 30 min, the prediction is significantly better than the daily profile. However, this time horizon is considerably shortened if there are trading events, such as the full hour dispatches.  

\subsection{Impact of the window size}
\label{sec:impact_window_size}

\begin{figure*}
    \centering
    \includegraphics[width=\textwidth]{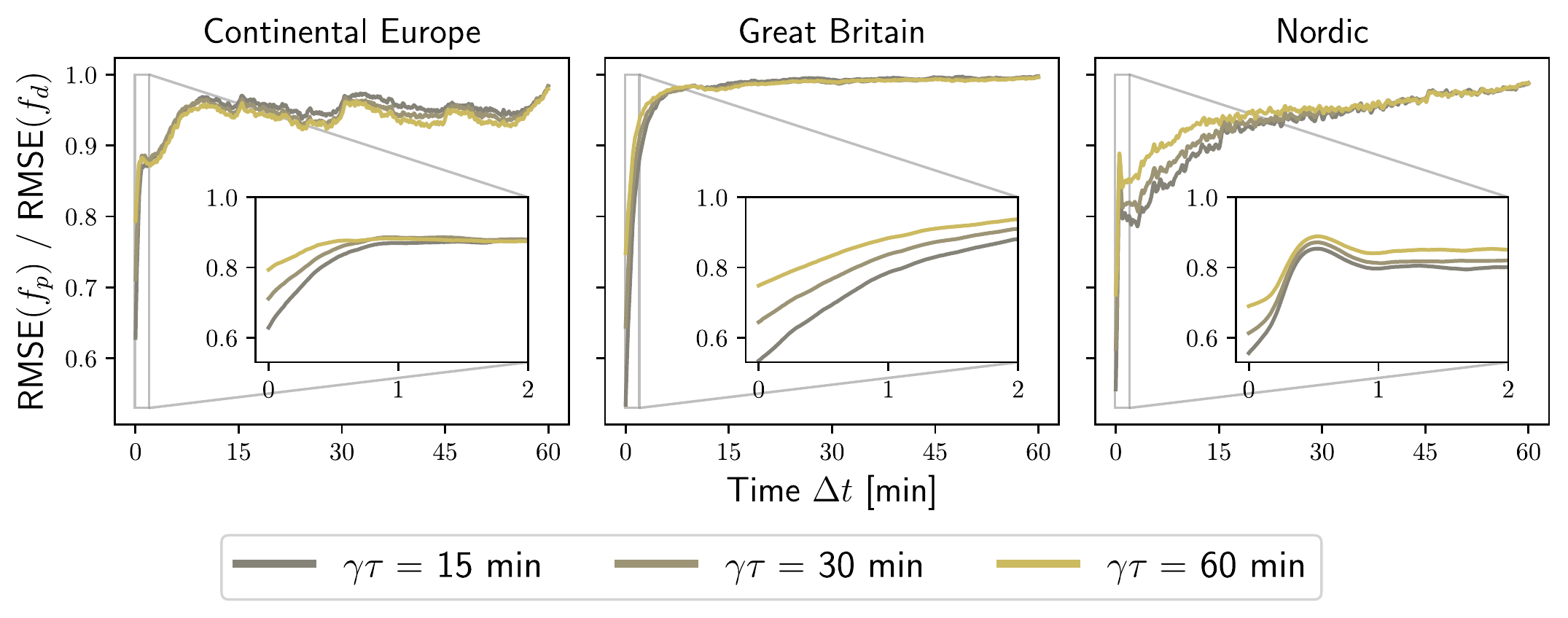}
    \caption{Shorter windows predict more accurately at the beginning. The optimal window size (or pattern length) $\gamma\tau$ is different depending on the prediction length. During the first minute, the shortest window performs best in all grid areas, as it contains more specific information about the near past. For several minutes to one hour, the results differ between the areas.}
    \label{fig:8_Past_window_size}
\end{figure*}{}

We finalize the discussion of the WNN predictor by shortly investigating the impact of different window sizes. In addition to the window size $\gamma\tau=60~$min (which we have used throughout this paper), we show the prediction errors for $\gamma\tau=15~$min and $30~$min in Fig. \ref{fig:8_Past_window_size}. 

On time scales of several minutes to one hour, there is no significant difference between the predictors in CE and GB. The large window is slightly better than the shorter ones. In contrast, the smallest window performs best in the Nordic area especially in the first 15 minutes. Shorter windows contain more specific information about the near past than longer windows. In the Nordic grid, the significance of very specific historic patterns thus prevail much longer than in the other grids. This is consistent with Section \ref{sec:start_time}, where we have seen that the memory of specific historic patterns reduces relatively slow in the Nordic area. 

On time scales below one minute the smallest window performs best for all grid areas (inset). Shortly after the prediction starts, the memory of the last few seconds determines the trajectory. Irrespective of the area, the shorter window thus performs best on this time scale, as it contains more specific information about about the near past of the trajectory. 

We conclude, that small window sizes are best for prediction horizons below one minute. For several minutes to one hour, large window sizes are slightly better in CE and GB. If computational resources are scarce, smaller window sizes can also be used here, as they are less computationally expensive but only slightly less accurate. In the Nordic area, small window sizes are the best even for several minutes. However, the  performance differences are small in all grid areas, which also justifies that we did not systematically determine the optimal value for $\gamma$, thus saving computational time during training.

\section{Discussion}
Summarizing, we have demonstrated how a $k$-nearest neighbor (WNN) approach provides an accurate forecast of the power grid frequency.  The predictor performs particularly well when using an adaptive number of nearest neighbors. 
Compared to previously existing forecasts of the power grid frequency \cite{bolzoni_real-time_2019,kaur_power_2013,dong_frequency_2014,bang_forecasting_2019}, we make three key contributions: First, we  introduce the daily profile as a relevant and system-specific null model.
Secondly, we offer an extensive statistical demonstration and discussion of the performance of the WNN method for different synchronous areas on time scales of several seconds to one hour. 
Thirdly, we interpret the time-dependent predictability and optimization results based on the economic and physical dynamics in the different synchronous areas. In that way, we establish machine learning techniques as valuable tools for an a posteriori assessment of power system operation and stability.

Our results can be used to improve power system stability. Since our estimates are more precise than the daily profile, they could be used to estimate necessary control power capacities. This is particularly interesting since we have a solid prediction horizon of about 60 minutes, making slower, typically cheaper forms of control available, instead of purely relying on expensive primary control \cite{Machowski2011, wood2013power}. Especially during the first 15-30 minutes, our predictor is significantly more accurate than the daily profile and could replace it for planning purposes.
Crucially, our analysis is not limited to any specific grid but can be applied to any power system, given sufficient data to train the algorithm.

We even gained valuable lessons when the predictor performed worst: The largest prediction errors are associated with unforeseen events that are also missed by the daily profile. Therefore, the introduced WNN predictor could also be used as a diagnostics tool to identify external perturbations,  where e.g. renewable generation \cite{golestaneh2016very} or singular demand patterns caused by large sports events \cite{chen2011analysis} impact the frequency dynamics.  Furthermore, even our worst predictions correctly returned the expected average and standard deviation of the frequency time series for the next hour. Hence, the predictor could be used as a worst-case estimator to determine how much control power will be maximally necessary during the next hour to guarantee stable operation.

Finally, we went beyond pure forecasting of the next sixty minutes of the power grid frequency dynamics but instead achieved a better understanding of the different synchronous area: Monitoring the number of nearest neighbors allowed us to distinguish deterministic and stochastic behavior of different synchronous areas but also of different time intervals. Our analysis reveals that before the market acts every 15 minutes, the time series becomes less predictable but becomes more predictable after the power has been dispatched. This insight could be used to modify dispatch strategies in order to minimize the unpredictable impact on the frequency, reducing the required control power and thereby saving money. 

Our results on the forecast of the power grid frequency can be extended in multiple directions in the future. Firstly, we were restricted by data availability. A similar forecast and interpretation could be developed and applied to power grid frequency time series from other regions in the world, e.g. data  from the Eastern Interconnection in the US or from the Irish grid, with its high wind penetration. Secondly, many alternative forecasting methods are available, from artificial neural networks (ANN) \cite{alpaydin2009introduction} and recurrent neural networks (RNN) \cite{haluszczynski2019good} to classical methods of time series prediction \cite{schelter_handbook_2006}. However, a fully comprehensive review of all available methods was beyond the scope of this study and will be left for the future. 
Finally, we are convinced that our approach to forecasting and machine learning as a tool to understand a system's dynamics should also be applied to other time series, such as renewable generation \cite{sharma2011predicting}, air pollution \cite{cogliani2001air,williams2020superstatistical} or the stock market \cite{zuo2012stock}.

\section*{Acknowledgments}
We thank Mark Thiele for fruitful discussions. Furthermore, we gratefully acknowledge support from the German Federal Ministry of
Education and Research (BMBF grant no. 03EK3055B) and the Helmholtz
Association (via the the \textit{Helmholtz School for Data Science in
Life, Earth and Energy} (HDS-LEE), the joint initiative \textit{Energy
System 2050 - A Contribution of the Research Field Energy} and via the
grant No. VH-NG-1025). This project has received funding from the
European Union’s Horizon 2020 research and innovation programme under
the Marie Sk\l{}odowska-Curie grant agreement No. 840825. 

\bibliographystyle{IEEEtran}
\bibliography{references}

\begin{IEEEbiography}[{
\includegraphics[width=1in,height=1.25in,clip,keepaspectratio]{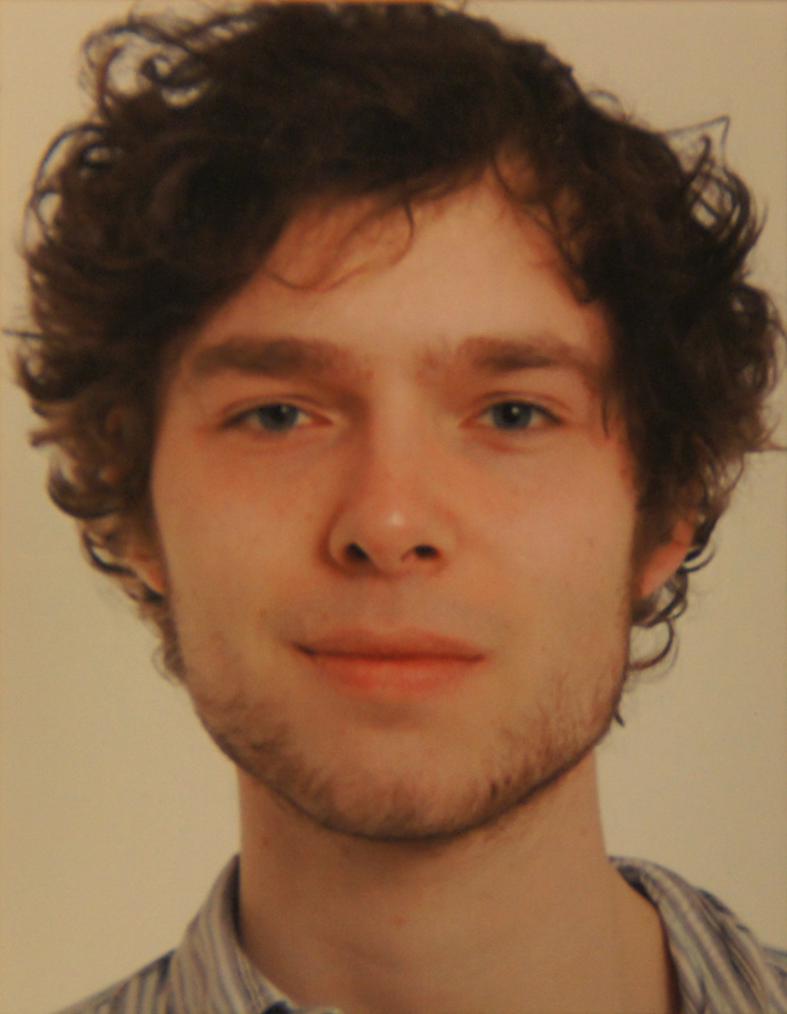}
}]{Johannes Kruse} received the B.Sc. and M.Sc. in physics at the Georg-August University of Göttingen, Germany in 2016 and 2019. During the final year of his studies, he wrote the Master's thesis at the Department of Engineering, Aarhus University, Denmark. Currently, he is pursuing a Ph.D. degree in the University of Cologne and the Forschungszentrum Jülich, Germany.
\end{IEEEbiography}

\begin{IEEEbiography}[{
\includegraphics[width=1in,height=1.25in,clip,keepaspectratio]{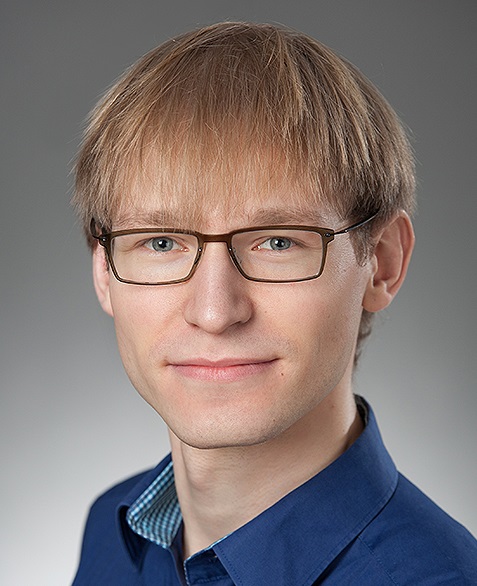}
}]{Benjamin Schäfer}received his Diplom degree in Physics from the University of Magdeburg, Germany in 2013. Pursuing his Ph.D. in Göttingen (Germany) London (United Kingdom) and Tokyo (Japan), he received his Ph.D. degree in physics in 2017 from the University of Göttingen. He has worked as a Postdoctoral Researcher at the Max Planck Institute for Dynamics and Self-Organization, Göttingen, Germany and the Technical University Dresden, Germany. Since 2019, he is working as a Marie Skłodowska-Curie Research Fellow at Queen Mary University of London.
\end{IEEEbiography}

\begin{IEEEbiography}[{
\includegraphics[width=1in,height=1.25in,clip,keepaspectratio]{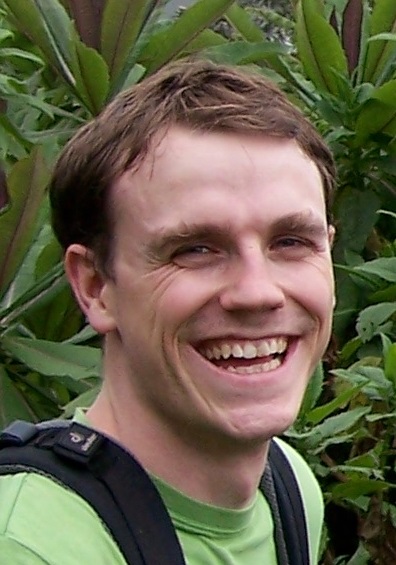}
}]{Dirk Witthaut}received the M.Sc. and Ph.D. degrees in physics from the Technical University of Kaiserslautern, Kaiserslautern, Germany, in 2004 and 2007, respectively. He has worked as a Postdoctoral Researcher at the Niels Bohr Institute, Copenhagen, Denmark and at the Max Planck Institute for Dynamics and Self-Organization, Gottingen, Germany and he has been a Guest Lecturer at the Kigali Institute for Science and Technology, Rwanda. Since 2014, he is leading a Research Group at
Forschungszentrum Julich, Germany and he is Assistant Professor at the University of Cologne.
\end{IEEEbiography}

\end{document}


\title{Supplementary Material: Data preparation}
\maketitle

\begin{figure*}
    \centering
    \includegraphics[width=0.8\textwidth]{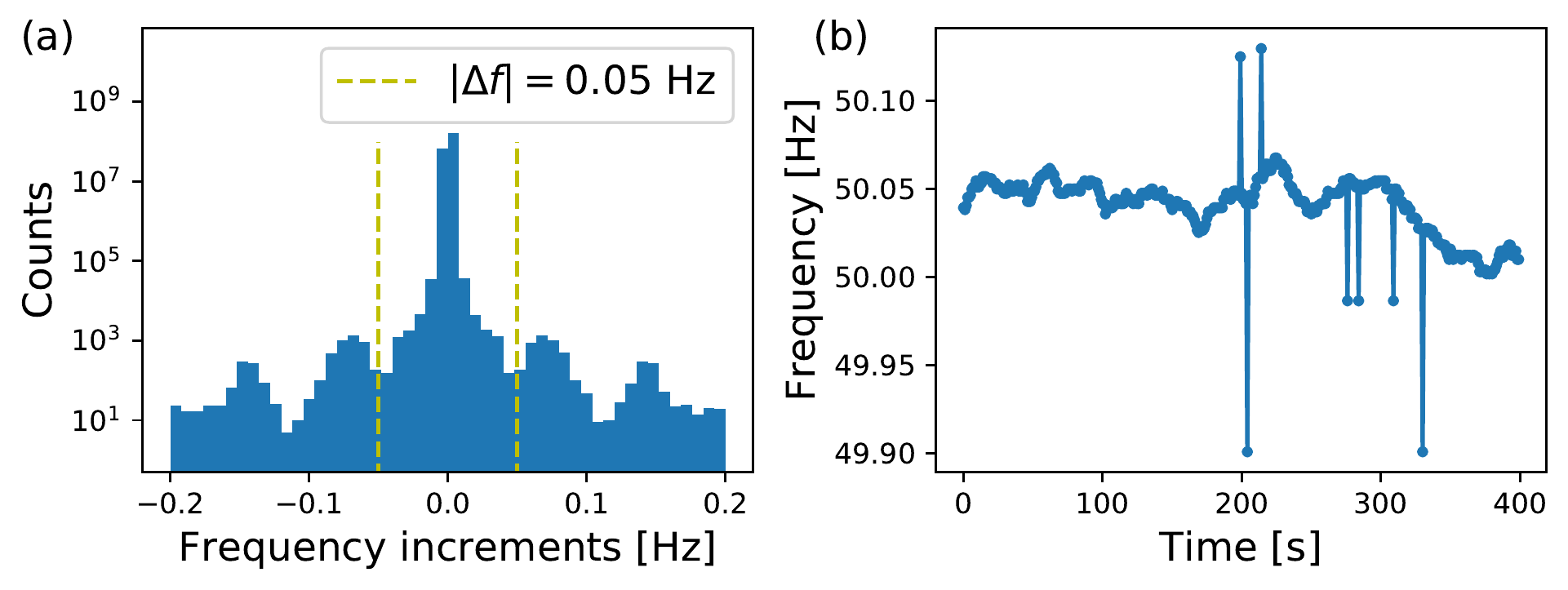}
    \caption{Isolated peaks exhibit increments larger than $\Delta f_c=0.05$ Hz. The Historgram (a) of the frequency increments in the CE data shows two minima for $\Delta f_c= \pm 0.05$ Hz. The frequency sample (b) from CE confirms that unrealistic isolated peaks exhibit frequency increments larger than 0.05 Hz. Both observations suggest the choice of $\Delta f_c=0.05$ Hz as a threshold for isolated peaks.}
    \label{fig:suppl_isolated_peaks}
\end{figure*}
\begin{figure*}
    \centering
    \includegraphics[width=\textwidth]{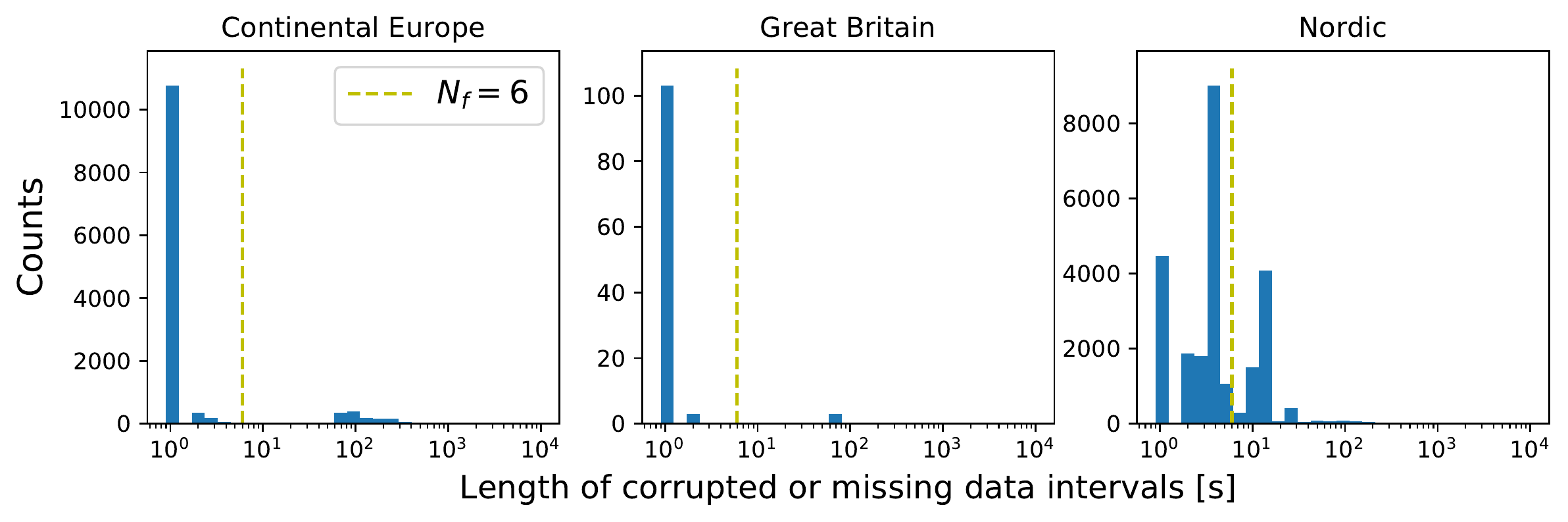}
    \caption{Missing and corrupted data points are marked as NaN-values. As a result, there are many intervals of NaN-values in the processed data. This figure displays the histogram for the lengths of these segments. CE and GB exhibit a particularly large amount of intervals with lengths below 6 time steps. Filling $N_f=6$ values thus eliminates many corrupted or missing data segments while not changing the data too much.}
    \label{fig:suppl_nan_intervals}
\end{figure*}

The frequency data for this paper has been pre-processed. A thorough pre-processing was necessary, as 
all datasets contain missing or corrupted data points and exhibit different formats and time resolutions. During the pre-processing we uniformly mark missing and corrupted data points, which offers the opportunity to individually fill, interpolate or exclude these invalid measurements. The whole process consists of two steps (Section \ref{app:processing_procedure}) and exhibits some free parameters that are fixed based on the data (Section \ref{app:parameter_choice}). Our code and the ready-to-use pre-processed data are available online \cite{data_repository}. 

Our time series include frequency measurements from the Continental Europe (CE), the Great Britain (GB) and the Nordic synchronous areas. The raw data is described in Table \ref{tab:data_details}. By the 18th February 2020, the data has been publicly available on the websites of different Transmission System Operators (TSO). We follow the naming convention established in \cite{noauthor_commission_2017} and use "area" synonymous with "synchronous area".

\begin{table*}
    \centering
    \caption{Raw frequency recordings are available from different TSOs. For each synchronous area, we obtained publicly available frequency data from one of the Transmission System Operators (TSO). TransnetBW is a German TSO, while Nationalgrid and Fingrid are British and Finish, respectively. Although we have only used the years 2015-2018 in this paper, we have pre-processed a longer period of time.}
    \label{tab:data_details}
    \begin{tabular}{|c|c|c|}
         TSO (synchronous area) & Time resolution [s] & Pre-processed period \\ \hline
         TransnetBW (Continental Europe) \cite{noauthor_regelenergie_nodate} & 1 & 2012-02-01 to 2019-08-31  \\ 
         Nationalgrid (Great Britain) \cite{noauthor_historic_nodate} & 1 & 2014-01-01 to 2019-07-01  \\
         Fingrid (Nordic) \cite{noauthor_frequency_nodate} & 0.1 & 2015-01-01 to 2019-11-30  
    \end{tabular}
\end{table*}

\section{Pre-processing procedure}
\label{app:processing_procedure}

\subsection{Convert data to common format} 
The data contains time stamps with varying formats and Daylight Saving Time (DST) changes. We thus process the time stamps and convert them to a uniform format. We point out that we use local time including the DST changes in the processed data. This is relevant for our weighted-nearest-neighbor prediction, since we compare frequency patterns based on their time stamp. By using the local time, we account for other important socio-economic patterns (such as the load) that rather follow local time than UTC time. In addition, we convert the time series to a common resolution of one second. We therefore resample the Nordic data by averaging it in non-overlapping windows of 10 data points. Finally, we mark missing time stamps by inserting a NaN-value (Not a Number) into the time series.

\subsection{Mark and clean corrupted data}
We search for different types of corrupted data. In particular, we identify isolated peaks, too high or low frequency values and too long windows with constant frequency. The identification is based on the frequency $f(t)$ or their increments $\Delta f(t) = f(t) - f(t-1)$ and follows these definitions:
\begin{itemize}
    \item A time stamp $t$ contains an isolated peak, if $\Delta f_t$ and $\Delta f_{t+1}$ have the opposite sign and are too large, i.e. $|\Delta f_t|, |\Delta f_{t+1}| > \Delta f_c $.
    \item Following \cite{Schaefer2017a}, frequencies below 49Hz and above 51Hz are consider as unrealistic and thus as too low (and too high). 
    \item We define a too long constant window as a time intervals of length $T > T_c$ that exhibits increments $|\Delta f| < 10^{-9}$ Hz.
\end{itemize}
These three types of data points are marked and converted to NaN-values. This offers the possibility to apply custom cleaning methods to the data (for example interpolation or data exclusion). Here, we clean the data by filling (at maximum) $N_f$ NaN-values with the last valid frequency value. During the prediction, we exclude frequency patterns with remaining NaN-values from the simulations. 

\section{Choice of pre-processing parameters}
\label{app:parameter_choice}
The above procedure exhibits three free parameters. We select the parameters $\Delta f_c$, $T_c$ and $N_f$ in the following way.
\begin{itemize}
    \item Isolated peaks mainly occur in CE data (Table \ref{tab:marked_data}). The increment histogram for CE (Fig. \ref{fig:suppl_isolated_peaks}(a)) shows minima at $\Delta f = \pm 0.05$ and separated maxima beyond this values. That indicates that the increments beyond this threshold are caused by another process than the regular stochastic frequency movement. The frequency sample in \ref{fig:suppl_isolated_peaks}(b) also shows that $\Delta f_c = 0.05$ would include most of the isolated peak. The GB and Nordic area do not show as many isolated peaks in their data  (Table \ref{tab:marked_data}). A manual inspection of the isolated peaks with $\Delta f_c = 0.05$ reveals, that no regular data points are accidentally marked as corrupted in the GB and Nordic data. We therefore keep the choice $\Delta f_c = 0.05$.
    \item We allow intervals of constant values below a length of $T_c=1$ min. We mainly predict the frequency on time intervals of several minutes, so that constant windows with $T<1$ min are negligible.
    \item We choose $N_f= 6$ since most of the corrupted CE and GB data can be cleaned in that way (Fig. \ref{fig:suppl_nan_intervals}). At the same time, we do not manipulate the time series too much on time scales relevant for our prediction. Note, that a large part of the marked data in the Nordic time series will not be filled by $N_f=6$s. This is mainly due to the large amount of missing data within the Nordic frequency recordings (Table \ref{tab:marked_data}). 
\end{itemize}
Using the procedure outlined above, we obtain clean data to be used for training and validation, see also \cite{data_repository}, where the clean data can be downloaded.

\begin{table*}
    \centering
     \caption{The number of corrupted data points differs between the synchronous areas. We mark different types of corrupted data points according to the definition in Section \ref{app:processing_procedure} and the parameters from Section \ref{app:parameter_choice}. Note, that we apply this procedure after resampling the data. This table thus shows the number of corrupted and missing data points (or intervals) for the data \textit{after} step (a) in Section \ref{app:processing_procedure}.}
     \label{tab:marked_data}
    \begin{tabular}{c|c|c|c|c|c|}
         Synchronous area & Isolated peaks & Too high frequency & Too low frequency & Long constant windows & Missing data intervals \\ \hline
         CE & 12229 & 10 & 8630293 & 1438 & 21  \\
         GB & 6 & 0 & 0 & 3 & 0 \\
         Nordic & 109 & 0 & 0 & 75 & 25103  \\
    \end{tabular}
\end{table*}

\bibliographystyle{IEEEtran}
\bibliography{references}